\documentclass[%
  superscriptaddress,
  preprintnumbers,
  amsmath,amssymb,
  aps,
]{revtex4-1}
\usepackage{graphicx}
\usepackage[osf]{mathpazo}
\usepackage{newpxtext}
\usepackage{enumerate}
\usepackage[caption=false]{subfig}
\usepackage{color}
\usepackage{ulem}
\usepackage{tikz}
\usetikzlibrary{shapes.misc}
\tikzset{cross/.style={cross out, draw=black, minimum size=2*(#1-\pgflinewidth), inner sep=0pt, outer sep=0pt},
cross/.default={1pt}}

\definecolor{wren}{RGB}{106,90,205}
\begin{document}
\preprint{KEK-TH-2344}
\preprint{J-PARC-TH-0248}
\preprint{RIKEN-iTHEMS-Report-21}
\title{Near-threshold Spectrum from Uniformized Mittag-Leffler Expansion\\
-Pole Structure of $Z(3900)$-}
\date{\today}
\author{Wren A. Yamada}
\email{wren-phys@g.ecc.u-tokyo.ac.jp}
\affiliation{Department of Physics, Faculty of Science, University of Tokyo, 7-3-1 Hongo Bunkyo-ku Tokyo 113-0033, Japan}%
\affiliation{Theory Center, Institute of Particle and Nuclear Studies (IPNS), High Energy Accelerator Research Organization (KEK), 1-1 Oho, Tsukuba, Ibaraki, 205-0801, Japan}%

\author{Osamu Morimatsu}
\email{osamu.morimatsu@kek.jp}
\affiliation{Department of Physics, Faculty of Science, University of Tokyo, 7-3-1 Hongo Bunkyo-ku Tokyo 113-0033, Japan}%
\affiliation{Theory Center, Institute of Particle and Nuclear Studies (IPNS), High Energy Accelerator Research Organization (KEK), 1-1 Oho, Tsukuba, Ibaraki, 205-0801, Japan}%
\affiliation{Department of Particle and Nuclear Studies,
Graduate University for Advanced Studies (SOKENDAI),
1-1 Oho, Tsukuba, Ibaraki 305-0801, Japan}%

\author{Toru Sato}
\email{tsato@rcnp.osaka-u.ac.jp}
\affiliation{Research Center for Nuclear Physics (RCNP), Osaka University, Ibaraki, Osaka 567-0047, Japan}%

\author{Koichi Yazaki}
\email{koichiyzk@yahoo.co.jp}
\affiliation{RIKEN iTHEMS, Wako, Saitama 351-0198, Japan}%

\begin{abstract}
We demonstrate how S-matrix poles manifest themselves as the physical spectrum near the upper threshold in the context of the two-channel uniformized Mittag-Leffler expansion, an expression written as a sum of pole terms (Mittag-Leffler expansion) under an appropriate variable where the S-matrix is made single-valued (uniformization).
We show that the transition of the spectrum is continuous as a S-matrix pole moves across the boundaries of the complex energy Riemann sheets
and that the physical spectrum peaks at or near the upper threshold when the S-matrix pole is positioned sufficiently close to it on the uniformized plane.
There is no essential difference on which sheet the pole is positioned. What is important is the existence of a pole near the upper threshold and the distance between the pole and the physical region, not on which complex energy sheet the pole is positioned.
We also point out that when the pole is close to the upper threshold, the complex pole does not have the usual meaning of the resonance.
Neither the real part represents the peak energy, nor the imaginary part represents the half width.
\par
Subsequently, we try to understand the current status of $Z(3900)$ from the viewpoint of the uniformized Mittag-Leffler expansion reflecting in particular, Phys.\,Rev.\,Lett.\,{\bf 117}, 242001 (2016) in which they concluded that $Z(3900)$ is not a conventional resonance but a threshold cusp.
We point out that their results turn out to indicate the existence of S-matrix poles near the $\bar D D^*$ threshold, which is most likely the origin of the peak found in their calculation of the near-threshold spectrum.
In order to support our argument, we set up a separable potential model which shares common behavior of poles near the $\bar D D^*$ threshold to the above-mentioned reference and show in our model that the structures near the $\bar D D^*$ threshold are indeed caused by these near-threshold poles.
\end{abstract}
\maketitle
\section{Introduction}
In recent years, many resonance-like enhancements which are potential candidates of exotic hadrons, have been observed in hadronic spectra near the thresholds of hadronic channels \cite{Guo:2017jvc, karliner}. 
Typical examples are the $XYZ$-mesons, such as $X(3872)$, $Y(4260)$, $Z(3900)$ in the charmonium spectra \cite{10.1093/ptep/ptw045}. 
There has been much discussion regarding the origins of these threshold enhancements, whether they are of resonant origin, such as tetra-quark, hadron molecular, or gluonic excitation states,
or of kinematic origin, such as threshold cusps \cite{GRAVESMORRIS1967477}, or triangle singularities \cite{Guo:2019twa}.
Clarification of such issues could possibly shed light on the complex nature of the strong interaction governing the dynamics of hadrons, and is of crucial importance in hadron physics.
\par
For analyses of the hadronic spectra, model-independent expressions such as the Breit-Wigner formula \cite{Breit:1936zzb}, Flatt\'{e} formula or equivalently the two-channel Breit-Wigner formula \cite{FLATTE1976224}, or coupled-channel model calculations are most commonly used in practice. 
Though most of these practices are conducted under the parameterization of energy or channel-momentum, analysis in the coupled-channel uniformized parameterization, introduced by Ref.\,\cite{KATO1965130}, and followed by Ref.\,\cite{PhysRev.188.2319,FUJII1975179}, 
can be particularly useful to clarify the interplay
between the position of the pole on the unphysical sheets of the $S$-matrix Riemann surface and its contribution to physical quantities such as the cross section or invariant-mass distribution.
In terms of the uniformized parameterization, a new method, the uniformized Mittag-Leffler expansion, was proposed and successfully applied to $\Lambda(1405)$ in Ref.\,\cite{PhysRevC.102.055201,Yamada:2021cjo},
in which the S-matrix or the Green's function is expressed as a simple sum of not only bound poles but also resonant poles.
\par
The purpose of the present paper is two-folded.
The first is to show how near-threshold poles of the S-matrix manifest themselves in the near-threshold spectrum. 
The second is to show that $Z(3900)$ can be naturally understood as a contribution of a set of poles near the $\bar D D^*$ threshold.
In Sec.\,II, III, we will demonstrate the general behavior of contributions of near-threshold poles on various sheets of the Riemann surface, including contributions commonly regarded as \lq\lq threshold cusps", based on the formalism of the Uniformized Mittag-Leffler expansion presented in Ref.\,\cite{PhysRevC.102.055201,Yamada:2021cjo}. 
Then in the following section, (Sec.\,IV), we will discuss under the two-channel uniformization parameterization, a plausible interpreteation of the $Z(3900)$ enhancement, reflecting on the HAL QCD results obtained in Ref.\,\cite{PhysRevLett.117.242001,Ikeda_2018} combined with the pole symmetry condition of the $S$-matrix. 
We will also set up a separable potential model shraing common features as the HAL QCD spectrum, which support our interpretation.
%
\section{Uniformized Mittag-Leffler Expansion}
Let us consider the spectrum of coupled double-channel two-body systems with threshold energies, $\varepsilon_1 < \varepsilon_2$.
The following findings of our analysis are valid in general, and are independent of the details of the system.
\par
Observables of the two-body system such as scattering cross sections or invariant-mass distributions are given by the scattering $T$-matrix or the two-body Green's function, which we generically write as ${\mathcal A}$.
We assume that ${\mathcal A}$ has only singularities of poles and two unitarity cuts running from the thresholds, $\varepsilon_1$ and $\varepsilon_2$, when expressed as a function of the complex center-of-mass energy, $\sqrt{s}$.

Let us define a dimensionless parameter, $e$, by
\begin{align*}
e&=\frac{s-\varepsilon_1^2}{\varepsilon_2^2-\varepsilon_1^2},
\end{align*}
where $e=0$ and $e=1$ at the lower and upper threshold, respectively.
We call $e$, scaled \lq\lq energy", because $e$ is real as long as the center-of-mass energy, $\sqrt{s}$, is real and is a monotonically increasing function of $\sqrt{s}$. 
Then, following Ref.\,\cite{KATO1965130}, we introduce a complex dimensionless variable, $z$, by
\begin{align*}
 z = e^{1/2} + \left(e-1\right)^{1/2} = &
 \begin{cases}
 i\left(\sqrt{-e} + \sqrt{1-e}\right) & e < 0, \\
 \sqrt{e} + i\sqrt{1-e} & 0 < e < 1, \\
 \sqrt{e} + \sqrt{e-1} & e > 1,
\end{cases}
\end{align*}
where $e^{1/2}$ ($\left(e-1\right)^{1/2}$) is the scaled \lq\lq momentum" in the lower (upper) channel.  
As a function of $z$, $e = \left(z + z^{-1}\right)^2/4$ and $\mathcal A$ is single-valued. Such a procedure where a function is made single-valued by introducing an appropriate parameterization is called uniformization \cite{Newton:book}. 
In this paper, for simplicity, we mainly deal with total scattering cross sections or missing-mass distributions, which are given by the imaginary part of the scattering $T$-matrix or the two-body Green's function, \cite{FetterWalecka,BERTSCH1975125,MORIMATSU1994679}, ${\rm Im}\, {\mathcal A}$.
\par
\begin{figure}[!htb]
\centering
\includegraphics[width=0.45\linewidth]{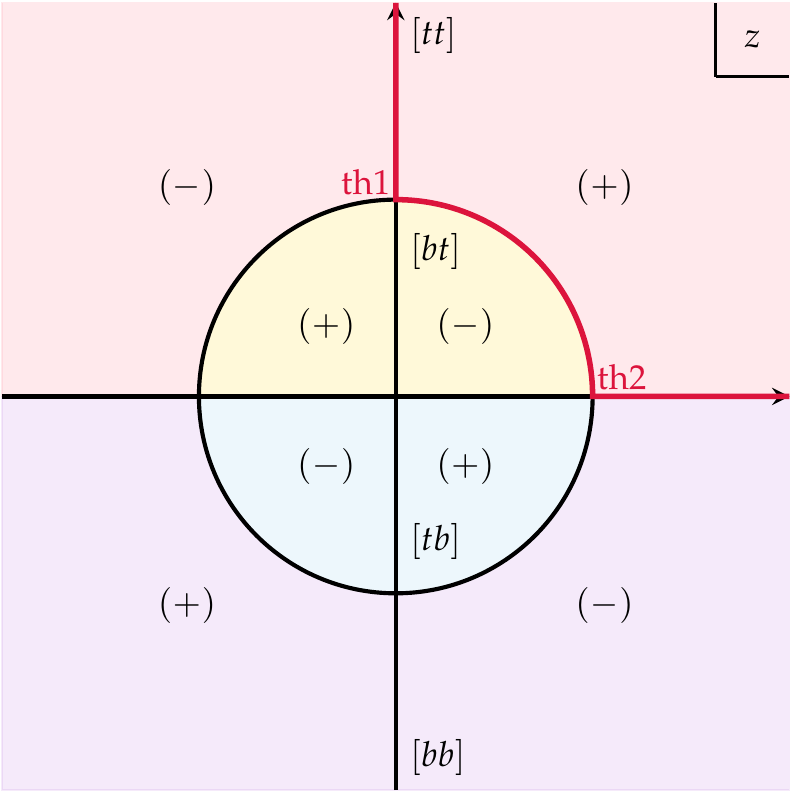}
\caption{The complex $z$ plane uniformizing the two-channel two-body S-matrix. The physical energy is represented by the red line. Below the lower threshold (labeled th1), the physical energy is on the imaginary axis. Between the lower and upper threshold (labeled th2), the physical region is on the unit circle, and then above the upper threshold, on the real axis.
The four domains specified by $[tt]$, $[bb]$, $[tb]$ and $[bt]$ correspond to four Riemann sheets of the complex energy
and $(+)$, $(-)$ show the imaginary part of the complex energy. For details refer to the main text.
}\label{fig:schem_z}
\end{figure}

The structure of the uniformized complex $z$ plane is shown in Fig.\,\ref{fig:schem_z}, where the physical energy is given by the red solid line.
We also show the domains corresponding to the four complex energy Riemann sheets labeled by the convention following Ref.\,\cite{PhysRevC.40.902}.
For each channel momentum, $k_i$, $t$ and $b$ refer to the top and bottom half of the complex momentum plane with the argument, $ 0 \le \arg k_i < \pi$ and $ \pi \le \arg k_i < 2\pi$, respectively.
Then, four complex energy Riemann sheets are specified by a set of complex channel momenta as [$tt$],  [$tb$], [$bt$] and [$bb$],
where e.g.\ [$tb$] means the momentum in lower (upper) channel is $t$ ($b$).
For later discussion we also show the regions where the imaginary part of the complex energy is positive and negative by $(+)$ and $(-)$, respectively.

In Ref.\,\cite{PhysRevC.102.055201,Yamada:2021cjo} it was shown that ${\mathcal A}$ can be expanded as a sum of pole terms in a form similar to that of Ref.\,\cite{HUMBLET1961529,ROMO197861,BANG1978381,BERGGREN1982261} in the variable, $z$, (uniformized Mittag-Leffler expansion) as,
\begin{align}
  {\mathcal A}(z) = -\frac{1}{\pi}\sum_{n}\left(\frac{r_n}{z-z_n}-\frac{r_n^\ast}{z+z_n^\ast}\right).\label{eq:unif_mle}
\end{align}
where $z_n$ and $r_n$ are, respectively, the position and residue of a pole in the variable, $z$.
We define a real dimensionless function, $f(z;z_p,\phi_p)$, by
\begin{align}
 f(z;z_p,\phi_p) = -\frac{1}{\pi}\text{Im}\,\left[\frac{\exp(i\phi_p)}{z-z_p}-\frac{\exp(-i\phi_p)}{z+z_p^\ast}\right].
\end{align}
Then, ${\rm Im}\,{\mathcal A}$ is given as
\begin{align}
  {\rm Im}\,{\mathcal A}(z) = \sum_{n} |r_n| f(z;z_n,\phi_n),
  \label{eq:uni.}
\end{align}
where $r_n=|r_n|\exp(i\phi_n)$. 
$f$ can be regarded as a 'normalized' pole-contribution linearly contributing to $\mathcal{A}$ by a weight of the absolute value of the residue.
\par
For later discussion, we give here a standard expression of the Breit-Wigner formula.
If the amplitude is analytic in the square root of the center-of-mass energy squared, $\sqrt{s}$, the normalized contribution from a pole near the physical region to the physical amplitude is given by
\begin{align*}
  f_{BW}(\sqrt{s}) & = -\frac{1}{\pi}{\rm Im}\,\frac{\exp(i\psi_p)}{\sqrt{s}-\sqrt{s}_p} \\
& = -\frac{1}{\pi}\frac{\cos\psi_p{\rm Im}\sqrt{s}_p+\sin\psi_p\left(\sqrt{s}-{\rm Re}\sqrt{s}_p\right)}{\left(\sqrt{s}-{\rm Re}\sqrt{s}_p\right)^2 + {\rm Im}\sqrt{s}_p^2}.
\end{align*}
Under the condition $\psi_p=0$, and the definition ${\rm Re}\sqrt{s}_p=\epsilon$ and ${\rm Im}\sqrt{s}_p=-\Gamma/2$, we obtain the standard Breit-Wigner form \cite{Breit:1936zzb},
\begin{align*}
  f_{BW}(\sqrt{s}) = \frac{1}{\pi}\frac{\Gamma/2}{\left(\sqrt{s}-\epsilon\right)^2+\left(\Gamma/2\right)^2}.
\end{align*}
\section{Pole contribution near the upper-threshold}
\begin{figure}[!htb]
\centering
\includegraphics[width=0.45\linewidth]{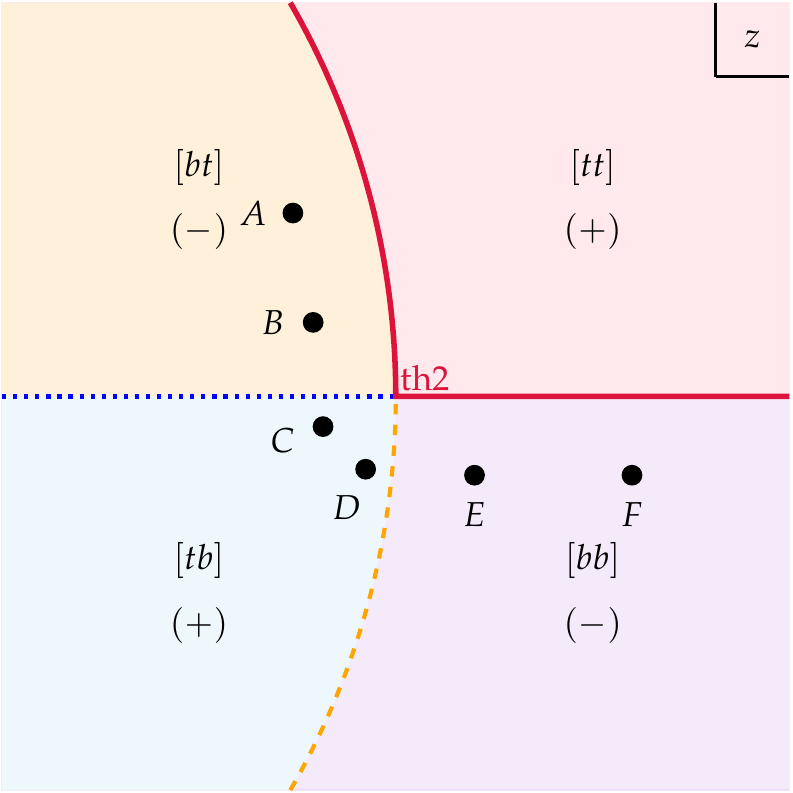}
\caption{Positions of the S-matrix poles, $A$-$F$, on the complex $z$ plane near the upper threshold (labeled th2).
The red line represents the physical region.}
\label{fig_2}
\end{figure}

In this section, we demonstrate how the near-threshold spectrum changes as a S-matrix pole moves from the [$bt$] sheet through the [$tb$] sheet to the [$bb$] sheet near the upper threshold.
We choose six positions, $A$-$F$, of the S-matrix pole near the upper threshold shown in Fig.\,\ref{fig_2}.
Their position on the complex $z$-plane, $z_p$, and their scaled energy, $e_p$, are tabulated in Table.\,\ref{table_1}.
We also show $e_0$, the scaled energy of the physical point nearest to the pole on the complex $z$ plane,
and $\gamma_0$, the distance between the pole and the nearest physical point, defined by
\begin{align*}
\begin{cases}
e_0 = \cos^2 \theta_p, \gamma_0 = 1-r_p & \text{[$bt$] sheet}\\
e_0 = 1, \gamma_0 = \sqrt{(x_p-1)^2 + y_p^2} & \text{[$tb$] sheet} \\
e_0 = \frac{1}{4}\left(x_p+\frac{1}{x_p}\right)^2, \gamma_0 = |y_p| & \text{[$bb$] sheet} 
\end{cases}
\end{align*}
where $r_p=|z_p|$, $\theta_p={\rm Arg}\, z_p$, $x_p={\rm Re}\, z_p$ and $y_p={\rm Im}\, z_p$.
\begin{table}[h]
\begin{tabular}{ccccccc}
\hline
\hline
&$A$&$B$&$C$&$D$&$E$&$F$\\
\hline
$z_p$&$0.869+0.233i$&$0.895+0.094i$&$0.908-0.038i$&$0.962-0.092i$&$1.100-0.100i$&$1.300-0.100i$\\
$e_p$&$0.943 - 0.053i$&$1.000 - 0.022i$&$1.007 + 0.008i$&$0.992 + 0.007i$&$1.002-0.018i$&$1.065-0.043i$\\
$e_0$&$0.933$&$0.989$&$1$&$1$&$1.01$&$1.065$\\
$\gamma_0$&$0.100$&$0.100$&$0.100$&$0.100$&$0.100$&$0.100$\\
sheet&[$bt$]&[$bt$]&[$tb$]&[$tb$]&[$bb$]&[$bb$]\\
\hline
\end{tabular}
\caption{The pole position on the complex $z$ plane, $z_p$, the complex pole energy, $e_p$, the nearest physical energy, $e_0$,
and the distance from the physical energy, $\gamma_0$, for poles $A$-$F$.}
\label{table_1}
\end{table}
\begin{figure}[!htb]
\centering
\includegraphics[width=\linewidth]{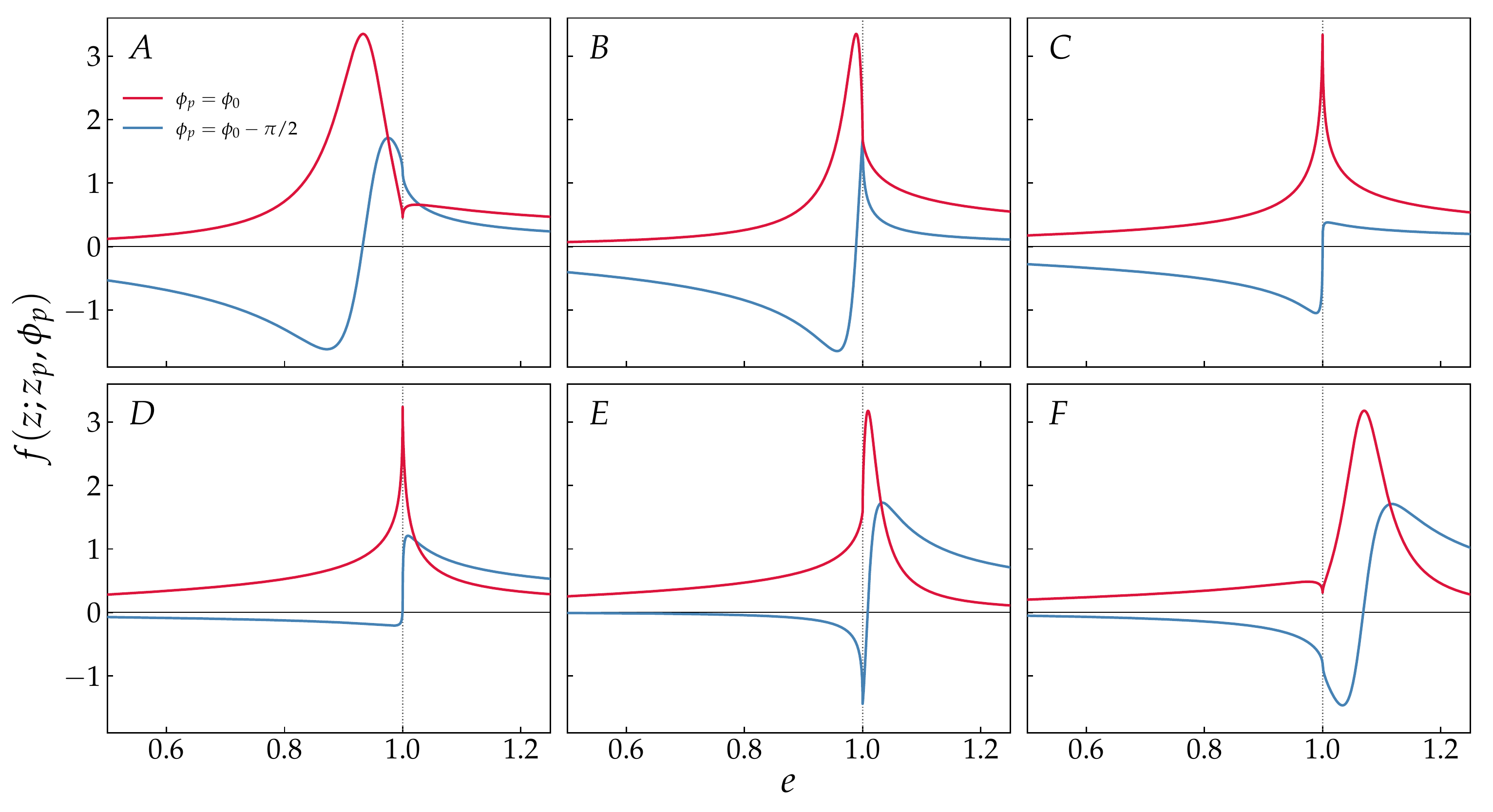}
\caption{The 'normalized' pole-pair contributions, $f(z;z_p,\phi_p)$, from poles $A$-$F$. Two cases are shown with different phases of residues, $\phi_p=\phi_0$, and $\phi_0-\pi/2$ where $\phi_0$ is respectively chosen as $\phi_0=\theta_p-\pi/2$, ${\rm{Arg}}(z_p-1)-\pi/2$, and $0$ for poles on the [$bt$], [$tb$], and [$bb$] sheet, so that the contribution, $f(z;z_p,\phi_p)$, is maximized at physical energy $e_0$.}\label{fig2}
\end{figure}
%
%
\par
The spectra of the contribution from poles, $A$-$F$, are shown respectively in Fig.\,\ref{fig2}.
As can be observed, when the pole is located on the [$bt$] sheet, the peak shows up below the upper threshold energy.
As the pole comes close to the [$tb$] sheet, the peak approaches the threshold.
Then, when the pole is on the [$tb$] sheet, the peak remains at the upper threshold while changing the slopes below and above the threshold.
Once the pole reaches the [$bb$] sheet, the peak position leaves the threshold, and increases again. 
Also, when the pole is close to the upper threshold, the complex pole energy does not have the usual meaning of the resonance.
Neither the real part represents the peak energy, nor the imaginary part represents the width.
Near the boundary of the [$bt$] and [$tb$] ([$tb$] and [$bb$]) sheets, ${\rm Re} \, e_p$ is larger (smaller) than 1 and ${\rm Im} \, e_p$ is very small,
while the position of the peak is below (above) or equal to 1 and the width of the peak remains almost unchanged.
Also, on the [$tb$] sheet ${\rm Im} \, e_p$ is positive.
The position of the peak is given by the energy of the physical point nearest to the pole on the complex $z$-plane,
and the width is given by the distance between the pole and the nearest physical point.
Finally, and most importantly, the transition of the spectrum is continuous.
There is no essential difference whether a pole is located on the [$bt$] sheet, [$tb$] sheet or [$bb$] sheet, contrary to the usual understanding that poles on the [$tb$] sheet are irrelevant.
Therefore, if the observed spectrum is peaked about the upper threshold with experimental uncertainties,
it would be difficult, in addition, non-essential to exactly determine which sheet the pole is really located. 
What is important is the existence of a pole near the upper threshold and how distant the pole is from the physical energy,
which should be sufficient for us to know  from experimental data.
\par
\begin{figure}[!htb]
\centering
\includegraphics[width=\linewidth]{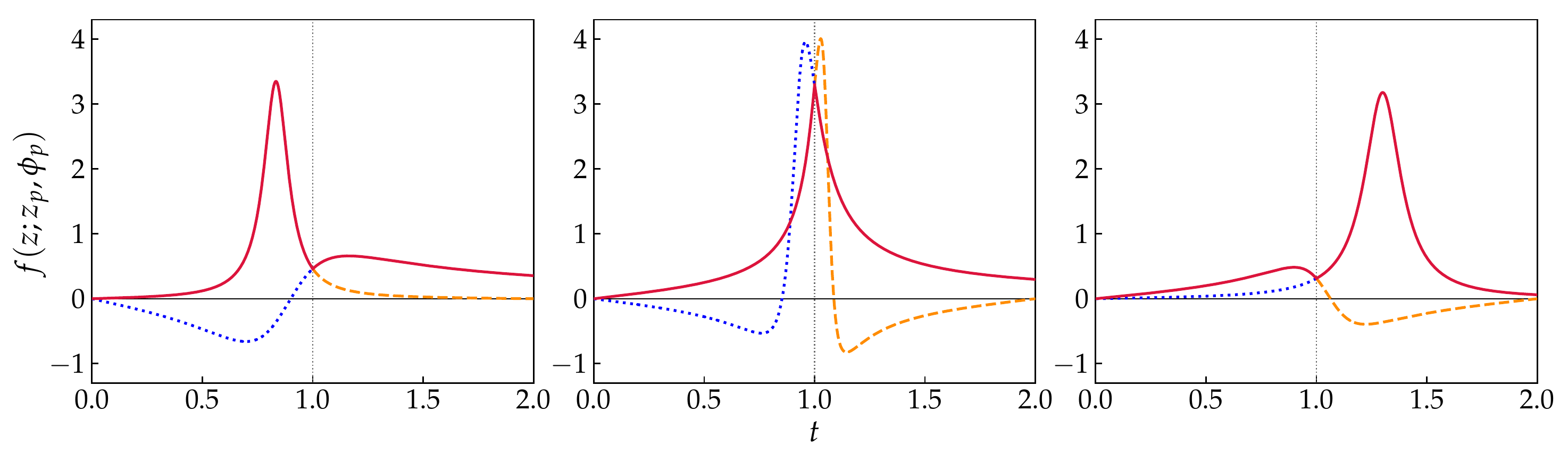}
\caption{
The behavior of $f(z;z_p,\phi_p)$ on the unit circle, $z=\exp i \frac{\pi}{2}(1-t)$ (the red solid line, $0 < t < 1$, and the orange dashed line, $1 < t < 2$), and on the real axis, $z=t$ (the blue dotted line, $0 < t < 1$, and the red sold line, $1 < t < 2$) for typical cases of poles on the [$bt$], [$tb$] and [$bb$] sheets.
The pole position on the $z$-plane is $0.869+0.233i$ ([$bt$] sheet), $0.929 - 0.071 i$ ([$tb$] sheet) and $1.300-0.100i$  ([$bb$] sheet), respectively.
}\label{fig_1}
\end{figure}

Let us discuss the general behavior of the contribution from a near-threshold pole in comparison to the Breit-Wigner form.
When $e>1$, $z$ is on the real axis and we parametrize $z$ and $z_p$ as $z=x$, $z_p=x_p+iy_p$, where $x$, $x_p$ and $y_p$ are real.
Then, a pole contribution is given by
\begin{align}
 -\frac{1}{\pi}\text{Im}\,\frac{\exp(i\phi_p)}{z-z_p} = -\frac{1}{\pi}\frac{-\cos\phi_p y_p +\sin\phi_p\left(x-x_p\right)}{\left(x-x_p \right)^2 + y_p^2},
 \label{Eq:real}
\end{align}
which is a Breit-Wigner form in the variable $x=z$ (real) with an additional phase.
If we take the phase of the pole as, $\phi_p=0$, the contribution is given by,
\begin{align*}
 -\frac{1}{\pi}\text{Im}\,\frac{\exp(i\phi_p)}{z-z_p} = \frac{1}{\pi}\frac{y_p}{\left(x-x_p \right)^2 + y_p^2},
\end{align*}
which is peaked at $x=x_p$.
Now let us consider the behavior
when $0<e<1$, i.e., $z$ is on the unit circle. We can parametrize $z$ and $z_p$ as $z=\exp(i\theta)$, $z_p=r_p\exp(i\theta_p)$,
where $\theta$, $r_p$ and $\theta_p$ are real parameters.
Then, a pole contribution becomes
\begin{align}
 -\frac{1}{\pi}\text{Im}\,\frac{\exp(i\phi_p)}{z-z_p} = -\frac{1}{\pi}\frac{\left(-r_p+\cos(\theta-\theta_p)\right)\sin(\phi_p-\theta_p)-\sin(\theta-\theta_p)\cos(\phi_p-\theta_p)}{4r_p\sin^2\frac{\theta-\theta_p}{2}+\left(1-r_p\right)^2},
\label{Eq:unit-circle}
\end{align}
similar to a Breit-Wigner form in the variable $\theta$ with a phase.
If we take $\phi_p=\theta_p-\pi/2$, then,
\begin{align}
 -\frac{1}{\pi}\text{Im}\,\frac{\exp(i\phi_p)}{z-z_p}
 = \frac{1}{\pi}\frac{1-r_p-2\sin^2\frac{\theta-\theta_p}{2}}{4r_p\sin^2\frac{\theta-\theta_p}{2}+\left(1-r_p\right)^2},
\end{align}
which is peaked at $\theta=\theta_p$.
Therefore, we can intuitively understand the behavior of $f(\sqrt{e}+\sqrt{e-1};z_p,\phi_p)$ in the whole energy region by combining Eq.\,({\ref{Eq:real}}) for $x>1$ and Eq.\,({\ref{Eq:unit-circle}}) for $0 < \theta < \pi/2$,
as shown in Fig.\,\ref{fig_1} for typical cases of poles on the [$bt$], [$tb$] and [$bb$] sheets.
As a typical pole on the [$bt$] ([$bb$]) sheet, we take $A$ ($F$), while on the [$tb$] sheet we take a pole between $C$ and $D$.
We have chosen the phase of the residue, $\phi_p$, such that the contribution is maximized at the physical energy nearest to the pole.  
\par
When the pole is
close to the physical region and distant from the thresholds, in the region of ($0 \ll \theta_p \ll \pi/2$) on the [$bt$] sheet or ($x_p \gg 1$) on the [$bb$] sheet, the contribution is of a Breit-Wigner form. The Breit-Wigner form along the unit circle with a peak at $\theta=\theta_p$ on the [$bt$] sheet and along the real axis with a peak at $x=x_p$ on [$bb$] sheet.
When the pole is on the [$tb$] sheet and is close to the upper threshold, two Breit-Wigner forms overlap.
Then, the contribution is given by connecting slopes of two Breit-Wigner forms and is peaked at the upper threshold, $x  = 1$ and $\theta = \pi/2$.

\section{$Z(3900)$ from the viewpoint of uniformized Mittag-Leffler expansion}\label{sec:z3900}
$Z(3900)$ is a candidate of exotic hadrons found by BESIII \cite{PhysRevLett.110.252001,PhysRevLett.112.022001,PhysRevD.92.092006,liu2015exotic} and Belle \cite{PhysRevLett.110.252002}, 
which appears as a peak in both the $\pi J/\psi$ and $\bar DD^*$ invariant mass spectra in the reaction, $e^+e^-\to Y(4260)\to\pi\pi J/\psi$ and $\pi\bar DD^*$,
and later confirmed by CLEO-c \cite{XIAO2013366}.
Our ultimate goal is to analyze the experimental data for $Z(3900)$ by means of uniformized Mittag-Leffler expansion and draw a model independent conclusion on the pole structure of $Z(3900)$ properties, such as the mass and the width or on which sheet the S-matrix pole of $Z(3900)$ is located.
Before carrying out the entire program, however, in this paper we confine our argument to addressing the possible origin of the $Z(3900)$ enhancement from the viewpoint of the uniformized Mittag-Leffler expansion.
\par
A symmetry condition of S-matrix poles plays a vital role in the following argument.
Let $S(\{k_i\})$ be the multi-channel S-matrix with $\{k_i\}$ as the set of channel momentum in all channels.
Then, due to unitarity, the S-matrix has the symmetry,
\begin{align}
  S(\{-k_i^*\}) = S^*(\{k_i\}).\label{eq:sym_condition}
\end{align}
From this relation, one can immediately conclude that if $\{k_i\}$ is a pole of S-matrix, so is $\{-k_i^*\}$, which will be referred to as the 'conjugate pole' in the following. 
Then, if the complex scaled energy $e$ is a pole of S-matrix, so is $e^*$ on the same Riemann sheet specified by the set of complex channel momenta.
In terms of the two-channel uniformization variable, $z$, the above symmetry is expressed as
\begin{align*}
  S(-z^*) = S^*(z)
\end{align*}
and the conjugate pole of $z$ is $-z^*$ on the same Riemann sheet symmetric about the imaginary axis.
\par
In subsection A, we reexamine the HAL QCD results including the poles conjugate to those given in Ref.\,\cite{PhysRevLett.117.242001} from the symmetry of the S-matrix
and calculating the pole contributions in the framework of the two-channel uniformized Mittag-Leffler expansion.
There, we cannot determine the relative contributions of the poles in the spectrum since we do not know the residues of the poles in Ref.\,\cite{PhysRevLett.117.242001}.
Then, in subsection B, we set up a separable potential model in which we have a pole near the $\bar D D^*$ threshold similar to Ref.\,\cite{PhysRevLett.117.242001}
and show that the contribution of such a pole really dominates the spectrum in the vicinity of the $\bar D D^*$ threshold.
\\[5pt]
\subsection{HAL QCD Results}
There have already been many theoretical studies, which try to clarify the structure of $Z(3900)$.
Among them we focus on the work by the HAL QCD collaboration\cite{PhysRevLett.117.242001,Ikeda_2018}.
They studied the $\pi J/\psi$-$\rho \eta_c$-$\bar D D^*$ coupled-channel interactions using (2+1)-flavor full QCD gauge configurations in order to study the structure of $Z(3900)$.
They also examined the pole positions of the S matrix on the complex energy plane focusing on those corresponding to usual resonances.
They found some poles located far from the physical region.
From this observation, they concluded that $Z(3900)$ is not a usual resonance but a threshold cusp.

We found it hard to understand their conclusion from the viewpoint of uniformized Mittag-Leffler expansion, in which the physical spectrum is given as a sum of pole contributions in terms of the uniformization variable.
In the following we would like to point out that their results do indicate the existence of the S-matrix pole near the $\bar DD^*$ threshold, which is most likely the origin of the peak found in their calculation.

\begin{figure}[!htb]
\centering
\subfloat{\includegraphics[width=0.45\linewidth]{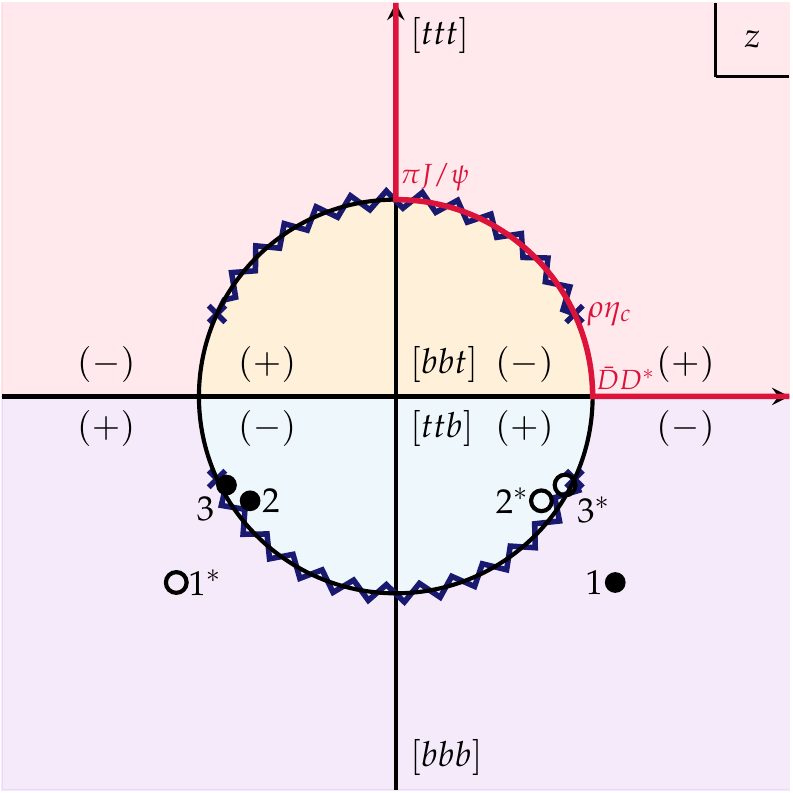}}
\hspace{10pt}
\subfloat{\includegraphics[width=0.45\linewidth]{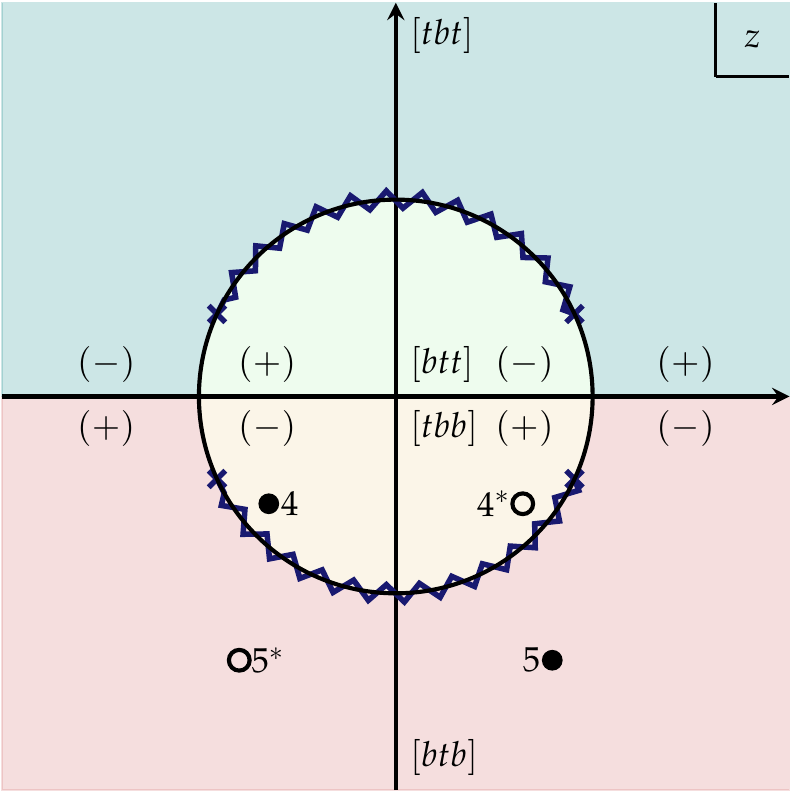}}
\caption{The (locally) uniformized complex $z$-plane for the $\pi J/\psi$-$\rho \eta_c$-$\bar D D^*$ coupled-channel S-matrix.
$\pi J/\psi$, $\rho\eta_c$, and $\bar D D^\ast$ denote the corresponding thresholds on the physical energy, respectively. The $z$-plane is a two-sheeted Riemann surface connected by the two branch cuts running along the unit circle.
Both the S-matrix poles given in Ref.\,\cite{PhysRevLett.117.242001}, $1$-$5$, and their conjugate poles $1^*$-$5^*$, (not given in Ref.\,\cite{PhysRevLett.117.242001}) are shown by filled and unfilled circles.}
\label{pole_s}
\end{figure}

In order to study the whole region of the $\pi J/\psi$-$\rho \eta_c$-$\bar D D^*$ coupled-channel, it would be ideal to implement three-channel uniformization, by which the three-channel S-matrix is single-valued on the whole plane of the unformization variable (global uniformization).
However, since the three-channel uniformization is very much involved, in this paper we focus on the region near the $\bar D D^*$ threshold.
The two-channel uniformization is sufficient for our purpose, by which the three-channel S-matrix can be regarded as single-valued near the $\bar D D^*$ threshold (local uniformization) but not on the whole plane of the unformization variable.
In the following, we employ the uniformization variable of $\pi J/\psi$-$\bar D D^*$ two-channel system. 
Effects of the coupling to the $\rho\eta_c$-channel will emerge as branch cuts in the complex $z$-plane as shown in Fig. \ref{pole_s}, which will be neglected.
\par
According to the pole symmetry condition, Eq.\,(\ref{eq:sym_condition}), there exist conjugate poles corresponding to the poles given in the HAL QCD results\,\cite{PhysRevLett.117.242001}.
Tab.\,\ref{hal_qcd_pole_pos} shows the scaled energy, $e_p$, the uniformizarion variables, $z_p$, for poles, $1$-$5$ (${\rm Im}\,e_p < 0$), given in Ref.\,\cite{PhysRevLett.117.242001} and for their conjugate poles, $1^*$-$5^*$ (${\rm Im}\,e_p > 0$),  not given in Ref.\,\cite{PhysRevLett.117.242001}.
\begin{table}[htbp]
\begin{tabular}{cccccc}
\hline\hline
&$1,1^*$&$2,2^*$&$3,3^*$&$4,4^*$&$5,5^*$\\
\hline
$z_p$&$\pm 1.11 - 0.95 i$&$\mp 0.74 - 0.53 i$&$\mp 0.86 - 0.45 i$&$\mp 0.65 - 0.54 i$&$\pm 0.79 - 1.34 i$\\
$e_p$&$0.60 \mp 0.41 i$&$0.66 \mp 0.09 i$&$ 0.79 \mp 0.02 i$&$0.60 \mp 0.17 i$&$0.16 \mp 0.44 i$\\
sheet&[$bbb$]&[$ttb$]&[$ttb$]&[$tbb$]&[$btb$]\\
\hline
\end{tabular}
\caption{The uniformization variables, $z_p$, and the scaled energy, $e_p$, for S-matrix poles, $1$-$5$ (${\rm Im}\,e_p < 0$), given in Ref.\,\cite{PhysRevLett.117.242001} and for their conjugate poles, $1^*$-$5^*$ (${\rm Im}\,e_p > 0$), not given in Ref.\,\cite{PhysRevLett.117.242001}. Also shown is the sheet on which each pole is positioned.}
\label{hal_qcd_pole_pos}
\end{table}
\begin{figure}[!htb]
\centering
\includegraphics[width=0.85\linewidth]{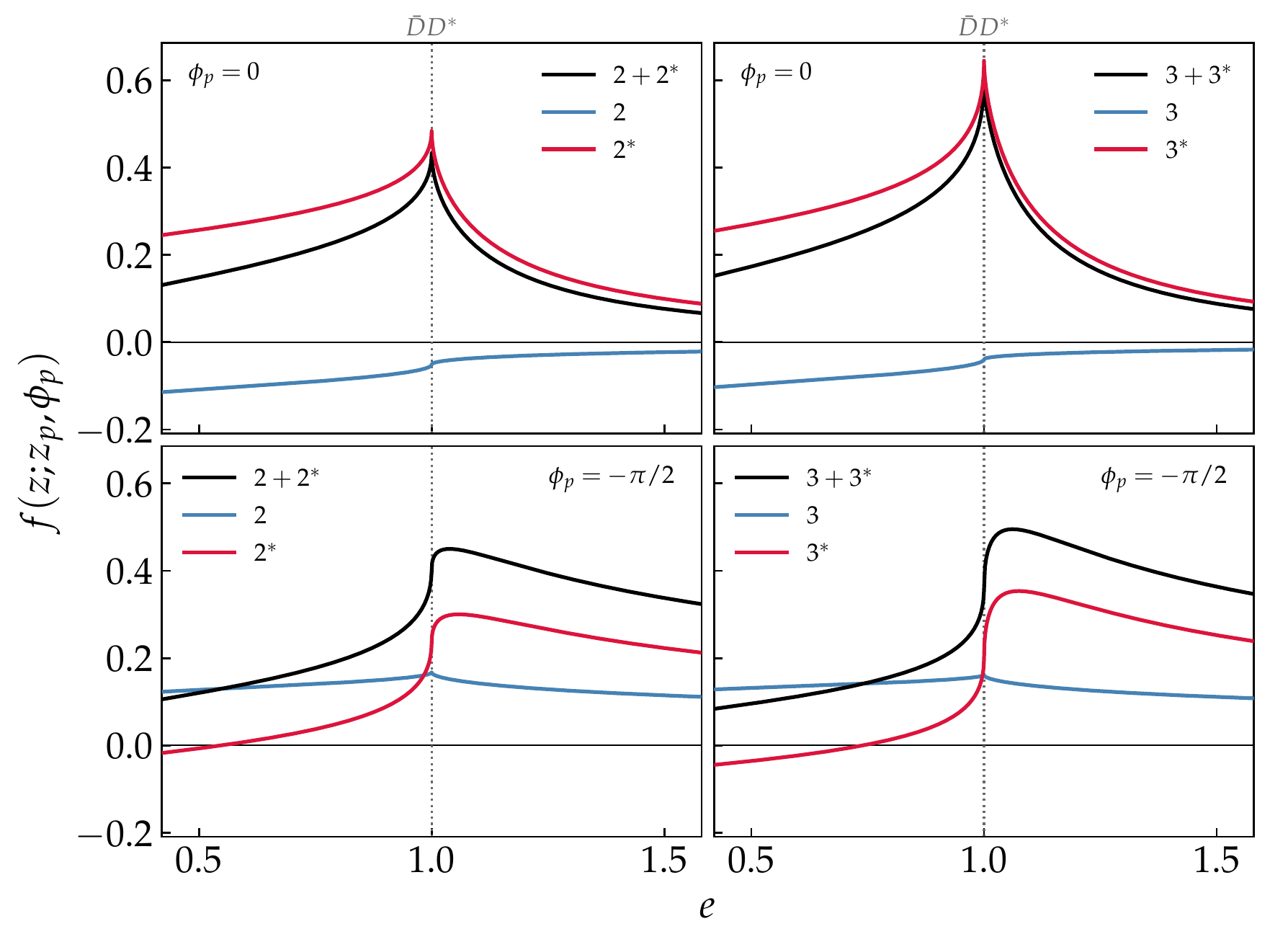}
\caption{'Normalized' pole-pair contributions, $f(z;z_p,\phi_p)$, from poles $2$, $2^*$ and $3$, $3^*$ for the cases where the phase of the residue is $\phi_p=0$ (above left, above right), and $\phi_p=-\pi/2$ (below left, below right).}
\label{spectrum_Z}
\end{figure}
Then, Fig.\,\ref{pole_s} shows where the poles are located on the complex $z$ plane. 
If one compares the location of poles $1$-$5$ and their conjugate poles $1^*$-$5^*$,
$1^*$ is much farther than $1$ from the physical region
but $2^*$ and $3^*$ are much nearer than $2$ and $3$ to the physical region.
In fact, among all the poles, $1$-$5$ and $1^*$-$5^*$, $3^*$ is the nearest to the physical region, the $\bar D D^*$ threshold, and the next is $2^*$.
Poles $4$ and $5$ together with their conjugate poles  $4^*$ and $5^*$ are located on the \lq\lq unphysical\rq\rq sheet of $z$ and are far from the physical region and therefore will not be discussed any more.
Fig.\,\ref{spectrum_Z} shows the normalized spectrum $f$ for pole-pairs, $2$-$2^*$ and $3$-$3^*$.
As expected, contributions from $2^*$ and $3^*$ are peaked at the $\bar D D^*$ threshold and their contributions are much larger than those of $2$ and $3$. 
\par
The above argument shows that the HAL QCD results indeed imply the existence of S-matrix poles near the $\bar DD^*$ threshold. The contribution from the near-threshold poles shows narrow structures at the $\bar DD^*$ threshold and is most likely the source of the observed $Z(3900)$ enhancement.
\subsection{Separable Potential Model}\label{sec:sep-pontential}
In this section we analyze the scattering amplitude of 
 a simple non-relativistic two-channel scattering.
We show an explicit case that a pole located
on the [$tb$] sheet plays an essential role for the enhancement of the scattering
amplitude at the upper threshold.
The scattering energy, $E$,
is expressed in terms of on-shell momentum $p_i \ (i=1,2)$
as,
\begin{eqnarray}
  E & = & \frac{p_i^2}{2\mu_i} + \Delta_i,
\end{eqnarray}
where $\mu_i$ and $\Delta_i$ are the reduced mass and sum of the masses
of two particles, respectively. We take $\Delta_2 > \Delta_1$. 
The interaction, $V_{ij}$,
is given by a separable form
\begin{eqnarray}
  V_{ij}(p',p) & = & g(p')v_{ij}g(p),
\end{eqnarray}
with a monopole form factor, $g(p) = \beta^2/(\beta^2 + p^2)$.
\par
We will keep only the off-diagonal potential, i.e., $v_{11}=v_{22}=0$ and $v_{12}=v$,
which is motivated from the effective potential for the
$\pi J/\psi$-$\rho\eta_c$-$D\bar{D}^*$ system given by the HAL-QCD results \cite{PhysRevLett.117.242001}.
The elastic scattering amplitude of the lower channel ${\cal F}_{11}$, which is related to the S-matrix as ${\cal F}_{11} = (S_{11} - 1)/(2i)$,
is given as,
\begin{eqnarray}
  {\cal F}_{11} & = & - \pi p_1\mu_1 g(p_1)^2 \frac{v^2 I_2}{1 - v^2 I_1I_2},
\label{eq-f}
\end{eqnarray}
with
\begin{eqnarray}
  I_i & = & \frac{\pi \mu_i \beta^3}{2 (p_i + i\beta)^2}.
\end{eqnarray}

To investigate the pole structure of the amplitude,
we rewrite Eq.~(\ref{eq-f}) using the uniformization variable, $z$,
\begin{eqnarray}
  z & = & \frac{p_1 + p_2 \sqrt{\frac{\mu_1}{\mu_2}}}{\Delta}
\end{eqnarray}
or
\begin{eqnarray}
  \frac{1}{z} & = & \frac{p_1 - p_2 \sqrt{\frac{\mu_1}{\mu_2}}}{\Delta}
\end{eqnarray}
with $\Delta^2 = 2\mu_1(\Delta_2 - \Delta_1)$.
The scattering amplitude is expressed in terms of two parameters
$\gamma$ and $\alpha$ as
\begin{eqnarray}
  {\cal F}_{11}(z) & = & - \alpha\gamma \frac{z^3(z^2+1)}{(z^2 - i\gamma z + 1)^2}
\nonumber \\
&&  \times \biggl[
  \frac{1}
{(z^2 + i\gamma z + 1)(z^2 + i\gamma' z - 1)  - \alpha z^2}
-
\frac{1}
{(z^2 + i\gamma z + 1)(z^2 + i\gamma' z - 1)   + \alpha z^2}\biggr],\label{eq:scta-amp}
\end{eqnarray}
where $\displaystyle  \gamma  =  \frac{2\beta}{\Delta}$,
$\displaystyle  \gamma'  =  \gamma \sqrt{\frac{\mu_1}{\mu_2}}$ and
$\displaystyle  \alpha  =  \frac{\pi\beta^3 v}{\Delta_2 - \Delta_1}$.
The poles of the amplitude are found by the following condition,
\begin{eqnarray}
z^4 -1 +(-\gamma\gamma' \mp \alpha) z^2 + i [(\gamma+\gamma')z^3 + (-\gamma+\gamma')z] = 0.
\label{eq-pole}
\end{eqnarray}
The coefficients of the even-power terms of $z$ are real,
the odd-power terms are pure imaginary and the constant term is $-1$.
Therefore, the eight solutions of Eq.~(\ref{eq-pole}) are given as $z_{p} = (z_1, -z_1^*, z_2, -z_2^*,ia_1,ia_2,ia_3,ia_4)$,
two complex poles $z_1,z_2$ together with their conjugate ones and four pure imaginary poles.
In addition, the amplitude has two poles of rank $2$ on the imaginary $z$ axis,
due to the form factor squared, $g(p)^2$.
Then, the scattering amplitude of Eq.\,(\ref{eq:scta-amp}) can be rewritten in the form,
\begin{eqnarray}
  {\cal F}_{11}(z) & = & 
  \sum_{j=1,2} \biggl[\frac{r_{j}}{z - z_{j}} + \frac{r_{j}^*}{z + z_{j}^*}\biggr]
  + \sum_{j=1,4} \frac{r'_j}{z - ia_j}
  +
  \sum_{j=1,2}\biggl[\frac{r_{\beta j}}{z - ia_{\beta j}}
      + \frac{i r'_{\beta j}}{(z - ia_{\beta j})^2}\biggr],
  \label{eq-ml}
\end{eqnarray}
where $a_j,~r'_j,~r_{\beta j},~a_{\beta j}$ and $r_{\beta j}'$ are real constants.
The first term includes complex poles and their conjugate ones.
Poles on the imaginary axis of $z$ appear in the second term.
The third term is due to the monopole form factor of the separable potential.
Eq.~(\ref{eq-ml}) explicitly shows that the amplitude is expressed by the uniformized Mittag-Leffler expansion as Eq.\,(\ref{eq:unif_mle}).
It is a nature of the separable interaction that the amplitude consists of a relatively small number of dynamical poles generated by repetition of the interaction and also of some poles due to interaction form factors.
In general, depending on the nature of the interaction or the model, the amplitude may contain singularities, such as poles, cuts or essential singularities, in addition to the dynamically generated poles.
Also, Eq.~(\ref{eq-ml}) explicitly shows that the amplitude satisfies the symmetry relation
\begin{eqnarray}
  {\cal F}_{11}(-z^*) = - {\cal F}_{11}^*(z).
\end{eqnarray}
Similarly, we show in Appendix that the Flatt\'e formula can be written in the form of the uniformized Mittag-Leffler expansion with 4 poles.

\begin{figure}[!htb]
  \centering
  \includegraphics[width=\linewidth]{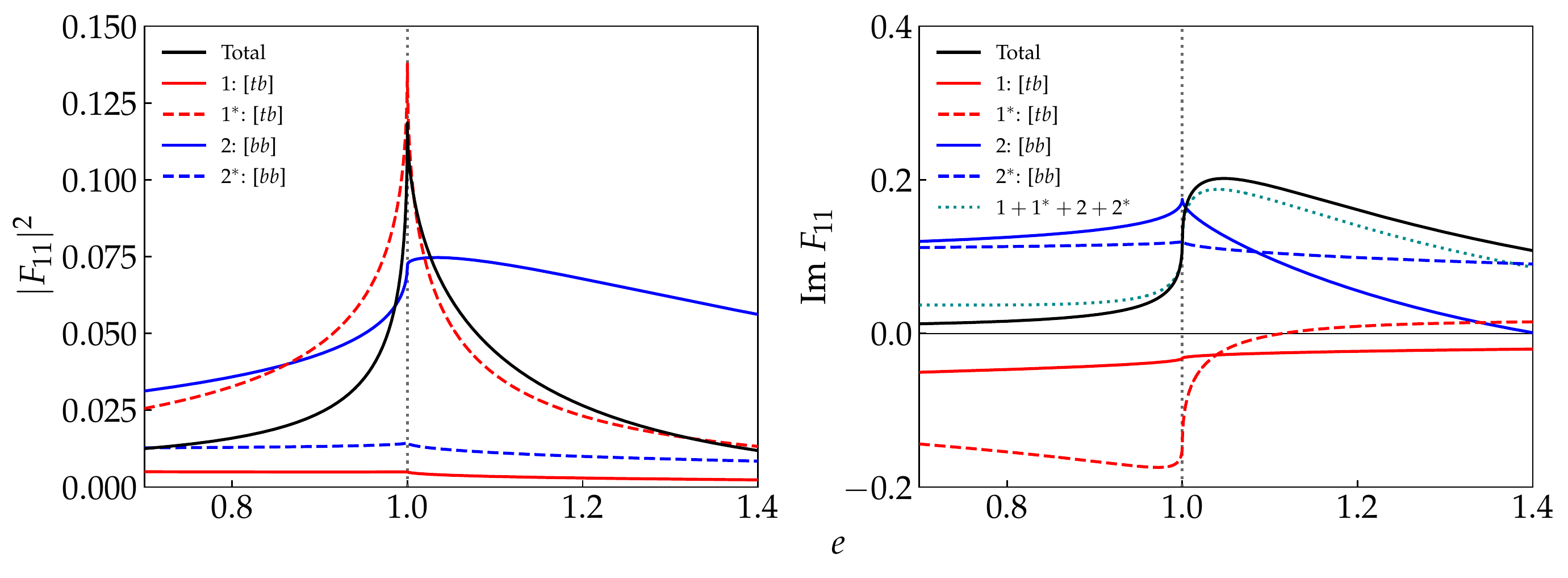}
  \caption{$|{\cal F}_{11}|^2$(left) and ${\rm Im}[{\cal F}_{11}]$(right) when $\alpha=\gamma=1.7$, $\sqrt{\mu_1/\mu_2} = 0.606$. $e$ is dimensionless parameter given by, $e=(E-\Delta_1)/(\Delta_2-\Delta_1)$.
Individual contributions of poles $z_i$ and their conjugate pairs are shown by solid and dashed curves (red/blue), respectively.
}\label{fig:fig-f}
\end{figure}

In the following numerical example, we take $\sqrt{\mu_1/\mu_2} = 0.606$,
where $\mu_1$ and $\mu_2$ are
the reduced mass of $\pi J/\psi$ and $\bar D D^*$, respectively.
By setting two parameters $\gamma = \alpha = 1.7$,
 poles emerge at
 $z = \pm0.754 - 0.229i$ on the [$tb$] sheet and  $z = \pm 1.198 - 1.076i$ on the
 [$bb$] sheet.
Tab.\,\ref{tab:num_result_seppoten} shows the uniformization variable, the residue and the scaled energy
for the poles.
It should be noted that the pole $1^*$ is located almost at the $\bar DD^*$ threshold energy with a small width.  
We examine $|{\cal F}_{11}|^2$ and ${\rm Im}\,{\cal F}_{11}$, which correspond to the elastic 
 cross section and the total cross section, respectively.
 In Fig.\,\ref{fig:fig-f}, black solid curves include contribution of
 all poles in Eq.\,(\ref{eq-ml}), while red (blue) solid and dashed
 curves show the contribution of pole $1$ ($2$) and its conjugate pole $1^*$ ($2^*$). 
The results clearly indicate that
the pole $1^*$ on the $[tb]$ sheet
is responsible for a significant fraction of the
 enhancement at the $\bar DD^*$ threshold.

\begin{table}[!htb]
  \begin{tabular}{ccc}
    \hline   \hline
    & $1,1^*$ & $2,2^*$\\
    \hline
    $z_p$        & $\mp 0.754 - 0.229i$      & $\pm 1.198 - 1.076 i$ \\
    $r_p$        & $ 0.118 \mp 0.042i$      & $-0.147 \mp 0.254i$ \\
    $e_p$        & $0.964 \mp 0.138i$      & $0.580 \mp 0.549 i$ \\
    sheet & [$tb$] & [$bb$] \\
    \hline    
  \end{tabular}
  \caption{Pole positions, $z_p$, and residues, $r_p$, of $F_{11}$ when $\alpha=\gamma=1.7$.}\label{tab:num_result_seppoten}
\end{table}
\par
The results from the separable potential model support our argument that the $Z(3900)$ enhancement can naturally be explained by the existence of poles near the $\bar DD^*$ threshold on the complex energy sheet which are usually considered to be physically irrelevant.
\section{Summary and conclusions}
In this paper we exhibited how S-matrix poles manifest themselves as the physical spectrum near the upper threshold, from the viewpoint of the uniformized Mittag-Leffler expansion, and argued that the enhancement of $Z$(3900) very close to the $\bar{D}D^\ast$-threshold most likely originates from a pole (a couple of poles) near the $\bar{D}D^\ast$-threshold.
\par
After introducing the two-channel uniformized plane and the uniformized Mittag-Leffler Expansion, we numerically demonstrated how the spectrum changes as a S-matrix pole moves near the upper threshold across the borders of complex-energy Riemann sheets.
As a pole moves from the [$bt$]-sheet, through the [$tb$]-sheet, to the [$bb$]-sheet, the contribution from the individual pole continuously transitioned, peaking at the energy of the physical point closest on the uniformized plane, with the width given by the corresponding distance.
The continuous behavior of the transition implies that the identity on which sheet a pole is located, is not essentially important.
The fundamental identities which characterize the contribution of a pole, is given by the residue of the pole, and the distance of the pole to the physical region on the uniformized plane.
We also observed that the complex pole energy does not have the usual meaning of the resonance when the pole is close to the upper threshold.
Neither the real part represents the peak energy, nor the imaginary part represents the half width. For example, the complex energy of a pole on the [$tb$] sheet in the near-threshold region of the upper threshold has a positive imaginary part.
  \par
Subsequently, we argued that $Z(3900)$ can be naturally understood as a contribution of a set of poles in the domain near the $\bar D D^*$ threshold.
We showed that the HAL QCD results \cite{PhysRevLett.117.242001,Ikeda_2018} combined with the symmetry argument of the S-matrix indicate the existence of the S-matrix poles near the $\bar D D^*$ threshold, whose normalized contributions have narrow structures at the threshold.
We further demonstrated such poles indeed exhibit a dominant enhancement in the spectrum by means of a separable two-channel non-relativistic potential model with a mono-polar form factor.
\par
To further solidify our claim, it would be very meaningful to develop a parameterization which enables us to globally (entirely) unformize the three-channel $S$-matrix. 
Also we would like to fully exploit our uniformized Mittag-Leffler method to actual experimental observations of $Z(3900)$, such as the invariant-mass distributions of $Y\to\pi\pi J/\psi$ measured by Belle, and BESIII collaborations
\cite{PhysRevLett.110.252001,PhysRevLett.112.022001,PhysRevD.92.092006,liu2015exotic,PhysRevLett.110.252002}.
Last but not least, we have not fully understood the nature of such a pole which is possibly responsible for the $Z(3900)$ enhancement and is located near the $\bar D D^*$ threshold on the complex energy sheet usually considered to be physically irrelevant.
It is not clear how we should interpret such a pole because it would have a complex energy with positive imaginary part and increases in time while a resonance pole has a complex energy with negative imaginary part and decreases in time.
It would be our future challenge to clarify the nature of such a pole, which would extend our understanding of the resonant phenomena in general.
\begin{acknowledgments}
The authors would like to thank Y. Ikeda for fruiteful discussions and for providing us information on the scattering amplitudes.
O. M. would like to thank the members of the discussion meeting held in KEK Tokai campus, Yoshinori Akaishi, Akinobu Dote, Toru Harada, Fuminori Sakuma, Shoji Shinmura and Yasuhiro Yamaguchi.
K. Y. is grateful to Quantum Hadron Physics group, iTHEMS at RIKEN and the director T. Hatsuda in particular for providing wonderful and stimulating working circumstances. 
This study of T. S. is in part supported by JSPS KAKENHI Grants No. JP19H05104.
\end{acknowledgments}
\appendix
\setcounter{secnumdepth}{0}
\section{APPENDIX: Flatt\'{e} form in the uniformized parameterization}
The Flatt\'e's formula \cite{FLATTE1976224} is widely used
to analyze the threshold enhancement and a possible role of the resonance.
The formula is essentially equivalent to the
two-channel Breit-Wigner (BW) formula \cite{PhysRevC.79.025205,PhysRev.188.2319,FUJII1975179}.
We rewrite the two-channel BW formula
in terms of the the uniformized Mitterg-Leffler expansion.
The elastic scattering amplitude of channel 1 is given as
\begin{eqnarray}
  {\cal F}_{11} & = & \frac{-\gamma_1 p_1}{E - M + i\gamma_1 p_1 + i\gamma_2 p_2}.
\end{eqnarray}
Here $M,\gamma_1,\gamma_2$ are parameters of real number and $p_1,p_2$ are
the momenta in the lower and upper channels defined in Sec.\,\ref{sec:sep-pontential}.
By using uniformization variable $z$, the amplitude is written as
\begin{eqnarray}
  {\cal F}_{11} & = &
  -\gamma \frac{z^3 + z}{z^4 + 1 + \alpha z^2 + i[(\gamma+\gamma')z^3  + (\gamma-\gamma')z]}.
\end{eqnarray}
where $\displaystyle \gamma = \frac{4\gamma_1\mu_1}{\Delta}$,
$\displaystyle \gamma'=\frac{4\gamma_2 \sqrt{\mu_1\mu_2}}{\Delta}$ and
$\displaystyle \alpha = \frac{4\mu_1}{\Delta^2}[\Delta_1 + \Delta_2 - 2M]$.
The denominator of the amplitude gives condition for the pole,
which is very similar to Eq.\,\ref{eq-pole}.
The two-channel BW amplitude has two pairs of poles $(z_1,-z_1^*),(z_2,-z_2^*)$
with $|z_1z_2|=1$.
The Flatt\'e formula always contains two pairs of poles and one is in the
[$tb$] or [$bt$] sheet and the other is in the [$bb$] sheet.
We can further rewrite the amplitude exactly as the
form of Mittag-Leffler expansion.
\begin{eqnarray}
  {\cal F}_{11} & = & \sum_{j=1,2} \biggl[ \frac{r_j}{z - z_j} + \frac{r_j^*}{z + z_j^*}\biggr]
\end{eqnarray}
The residue is constrained from unitarity. The S-matrix element is written as
\begin{eqnarray}
  S_{11} & = & \prod_{i=1,2} \frac{(z + \frac{1}{z_i})(z - \frac{1}{z_i^*})}
  {(z-z_i)(z+z_i^*)}.
\end{eqnarray}
\bibliography{paper_z3900}
\end{document}